# Quasar Redshifts: The Intrinsic Component

Peter M. Hansen, Revised 07 July 2016

Abstract:

*The large observed redshift of quasars has suggested large cosmological distances and a corresponding enormous energy output to explain the brightness or luminosity as seen at earth. Alternative or complementary sources of redshift have not been identified by the astronomical community. This study examines one possible source of additional redshift: an intrinsic component based on the plasma characteristics of high temperature and high electron density which are believed to be present.*

**Introduction**

An earlier study (Hansen 2014) derived intrinsic redshifts for a model of Active Galactic Nuclei (AGN) using a Monte Carlo large data simulation of correlations in the dielectric susceptibilities of the plasma in the broad line emission region (BLR). The AGN model used was based on published theories and estimates of the physical properties in that region. In brief, a set of "clouds" were used to simulate the dynamic, electromagnetic, and optical scattering occurring in the region that led to intrinsic redshifts.

There are no generally accepted physical models for the quasar or quasi-stellar object (QSO) environment. It seems to be clear, however, that the temperature and electron densities are greater than those present in an AGN. The properties of the ambient plasma should therefore be taken into account in the same manner and explored for the possibility of any intrinsic redshifts.

**Scaling from AGN to QSO Environments**

In Figure 1 the plasma region of AGN BLR temperatures and electron densities are shown in orders of magnitude that have appeared in many sources and models over several years (see Hansen 2014 and references therein). QSO temperatures and electron densities have also been estimated as shown in the figure (Elvis 2000 and Ruff 2012 respectively). However, no consistent models exist for the QSO's with the exception of rough physical estimates as selected and mentioned above. Some models do exist for AGN's and Seyfert's 1 & 2 (e.g. see Peterson 1997, Figure 7.1). It remains questionable, however, if these models represent QSO's in general.

The key parameter for the production of intrinsic redshifts arising from optical scattering in the plasma environment is the autocorrelation distance of the dielectric susceptibility (Wolf *et al* 1989a). Figure 2 shows, as a function of electron density, the autocorrelation distances that were derived from the AGN/BLR study mentioned above. These are extrapolated into the region of the QSO environment (Figure 1).

The Debye shielding distance or plasma "length" ($\sigma_D$), is taken as the upper limit to which plasma particles can "communicate" field effects or autocorrelate in dielectric susceptibility. It is given in terms of the electrical permittivity, Boltzmann constant, electron temperature, density, and charge ($\epsilon_0$, $k_B$, $T_e$, $n_e$, and $e$ respectively) as:

$$\sigma_D = \left(\frac{\epsilon_0 k_B T_e}{n_e e^2}\right)^{1/2} \quad (1)$$

The region, then, between the upper Debye limit and the lower limit or extrapolation of the AGN correlations as shown in Figure 2 is taken as the possible limits for the QSO plasma.

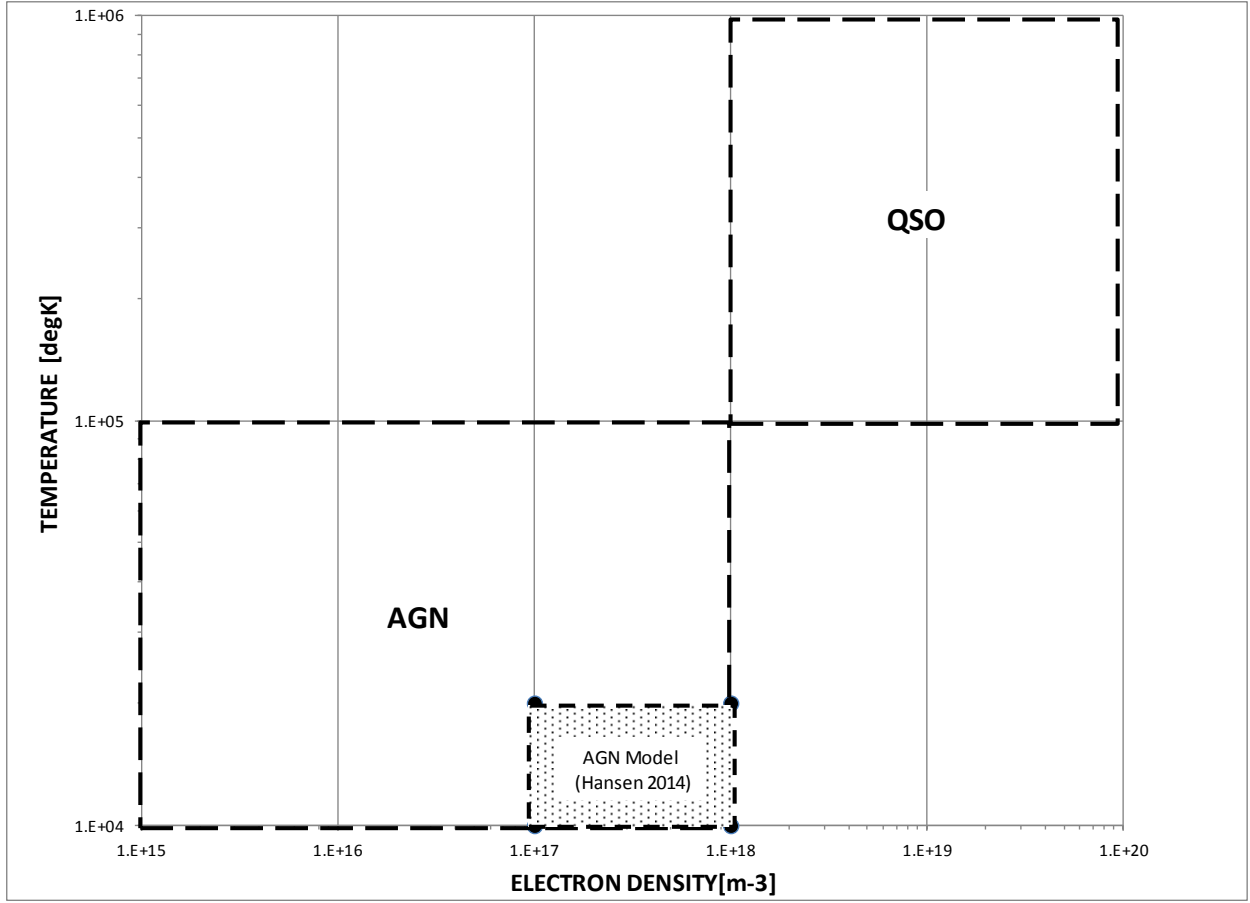

Figure 1. AGN and QSO Plasma Properties

The intrinsic plasma redshift ($z_2$) is given by Wolf (Wolf *et al* 1989a) as:

$$z_2(\Gamma_0, \lambda_0, \sigma, \gamma) = 4\left(\frac{\Gamma_0}{\lambda_0}\right)^2 \left[\left(\frac{2\pi\sigma}{\lambda_0}\right)^2 sin^2\left(\frac{\gamma}{2}\right) - 1\right] \quad (2)$$

Here, $\lambda_0$ is an optical emission line wavelength, $\Gamma_0$ is the equivalent line width, $\sigma$ is the plasma dielectric susceptibility correlation length, and $\gamma$ is the scattering angle assumed to be isotropic $(0 < \gamma < 2\pi)$.

In general, the wavelength or redshift ($z = \lambda - \lambda_0/\lambda_0 = \Delta\lambda/\lambda_0$) is frequency or wavelength dependent as is the plasma scattering medium; however, the magnitude of the shift is roughly on the order of, or smaller than, the equivalent line width (Wolf 1989b). This is illustrated in Figure 3 by using equation (2) as: $\Delta\lambda = z\lambda_0$ with the sine function maximized to unity for five typical emission lines and their equivalent width (Peterson 1997) using a representative and minimum correlation length in the QSO region ($\sigma = 2E-7$, see Figure 2).

It should be emphasized here that the Wolf correlation and scattering effects have been observed in both optical (Gori et al 1988) and acoustic (Bocko *et al* 1987) experiments

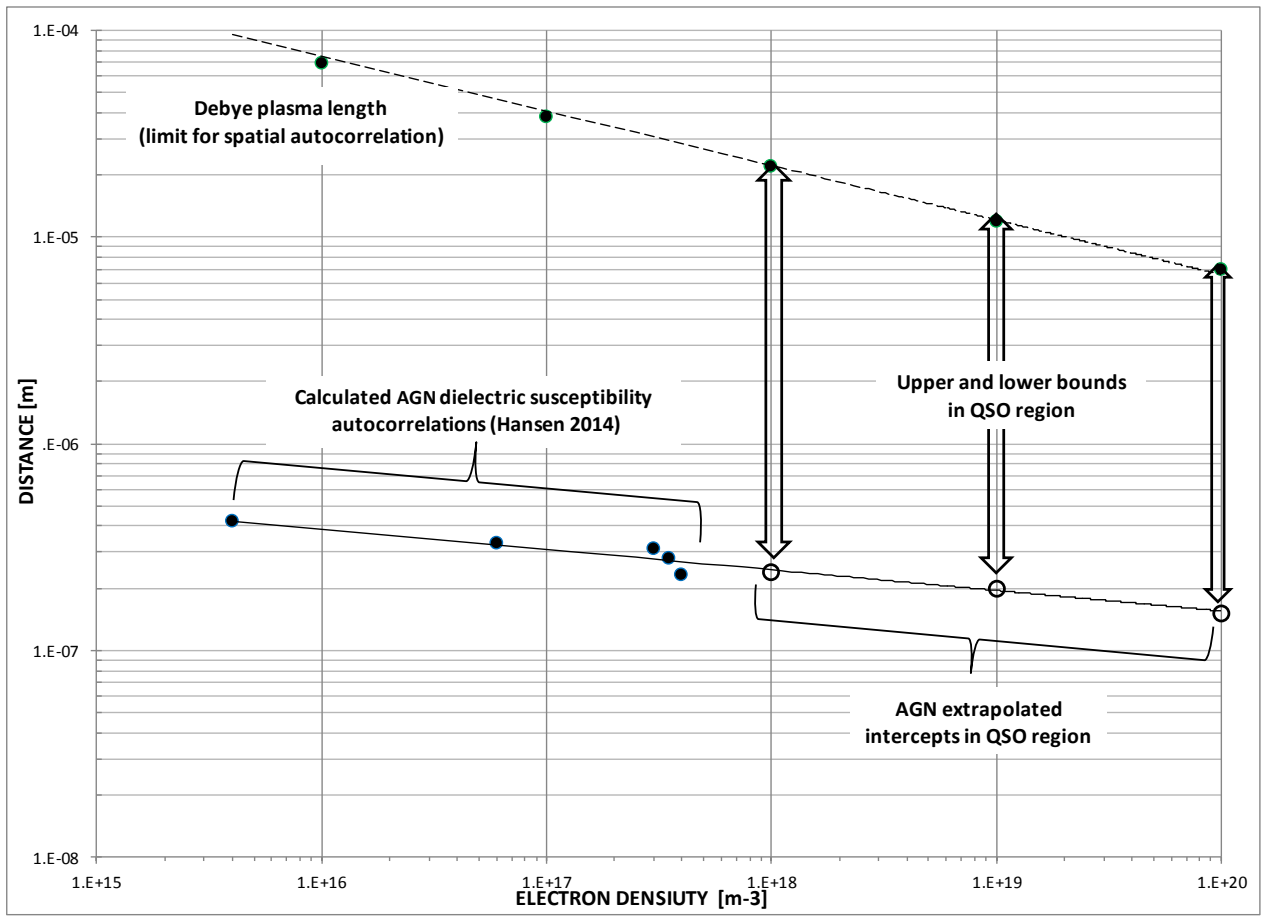

Figure 2. Autocorrelation Distances in AGN and QSO Plasma Regions

The intrinsic redshift ($z_2$) is a component of the total, or observed redshift ($z$) which also contains a recessional component ($z_1$) according to the relation:

$$(z + 1) = (z_1 + 1)(z_2 + 1) \qquad (3)$$

Since there are no data to determine where, between the upper and lower bounds, of the correlation lengths in Figure 2 the QSO value may physically be, a parametric plasma correlation length is defined in terms of a multiplier "$w$" as shown in Equation 4[1]:

$$\sigma = w\left(\frac{\lambda_0}{2\pi}\right), \ where: \sigma_{AGN} < \sigma < \sigma_D \qquad (4)$$

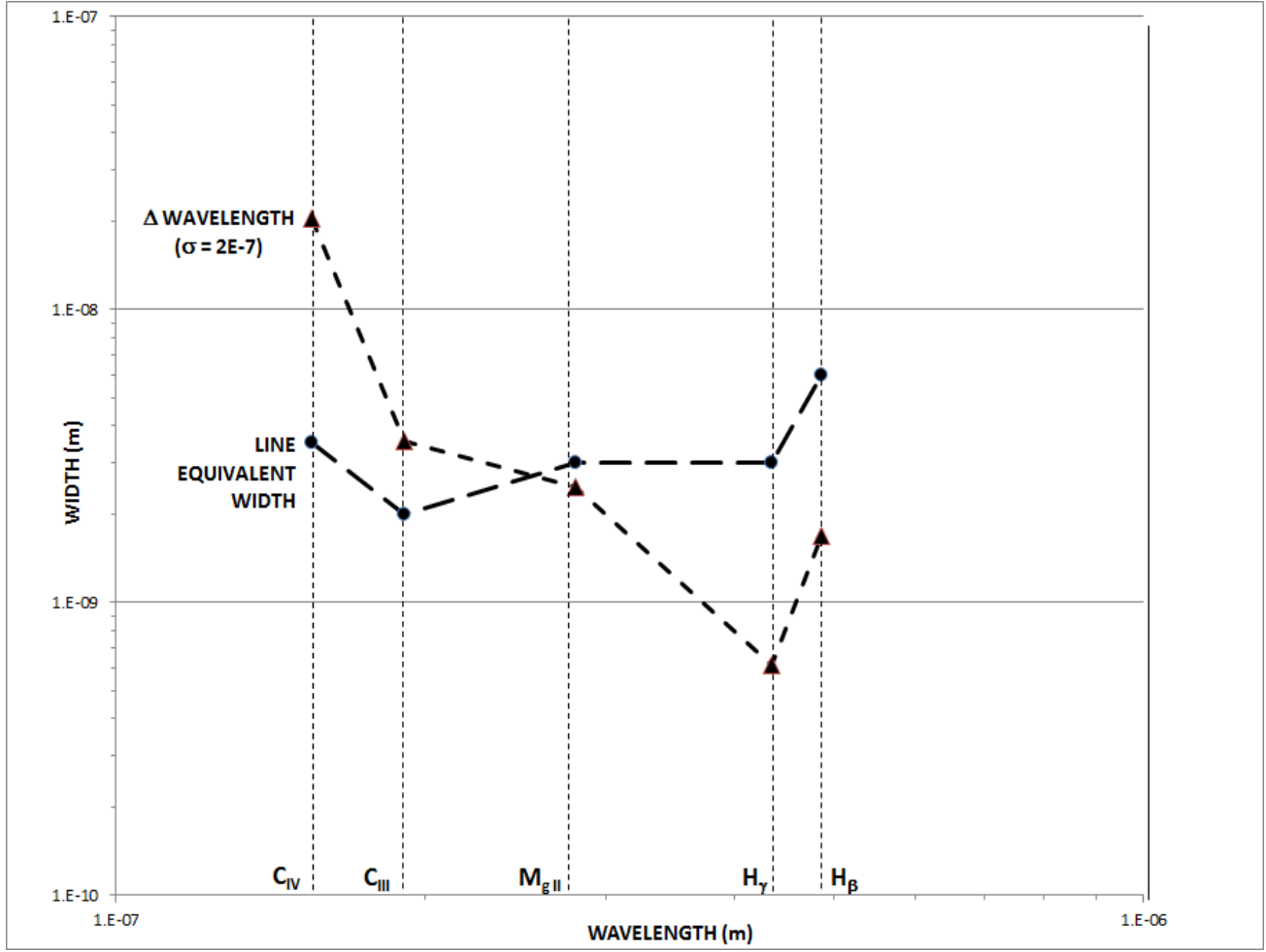

Figure 3. Emission Line Widths and Wavelength or Wolf Shifts

Given correlation distances and isotropic scattering, intrinsic redshifts can be calculated from equation 3 as shown in Figure 4. Note that $z_2$ rapidly becomes very small ($<<10^{-3}$) at low scattering angles and in fact becomes negative, or a blueshift, at scattering angles less than about ten degrees (relative to the line of sight between observer and object). Since no blueshifts are observed in QSO spectra the observed redshift is dominated by the recessional component and/or the total scattering especially from contributions at larger angles. The upper and lower QSO bounds shown correspond to those of Figure 2. As the correlation distance $\sigma(w)$ increases, so does the intrinsic redshift $z_2$ which is taken at the maximum scattering angle. Also shown in Figure 4 as a limiting upper bound is the maximum redshift ($z$) in a selected data base described below, and also for comparison one of the highest empirical redshifts observed in QSO's: ULAS J1120+0641.

---

[1] Here, $\lambda_0$ is taken as the $M_{g\,II}$ line at 2.798E-7 m (the same line example used in Hansen 2014).

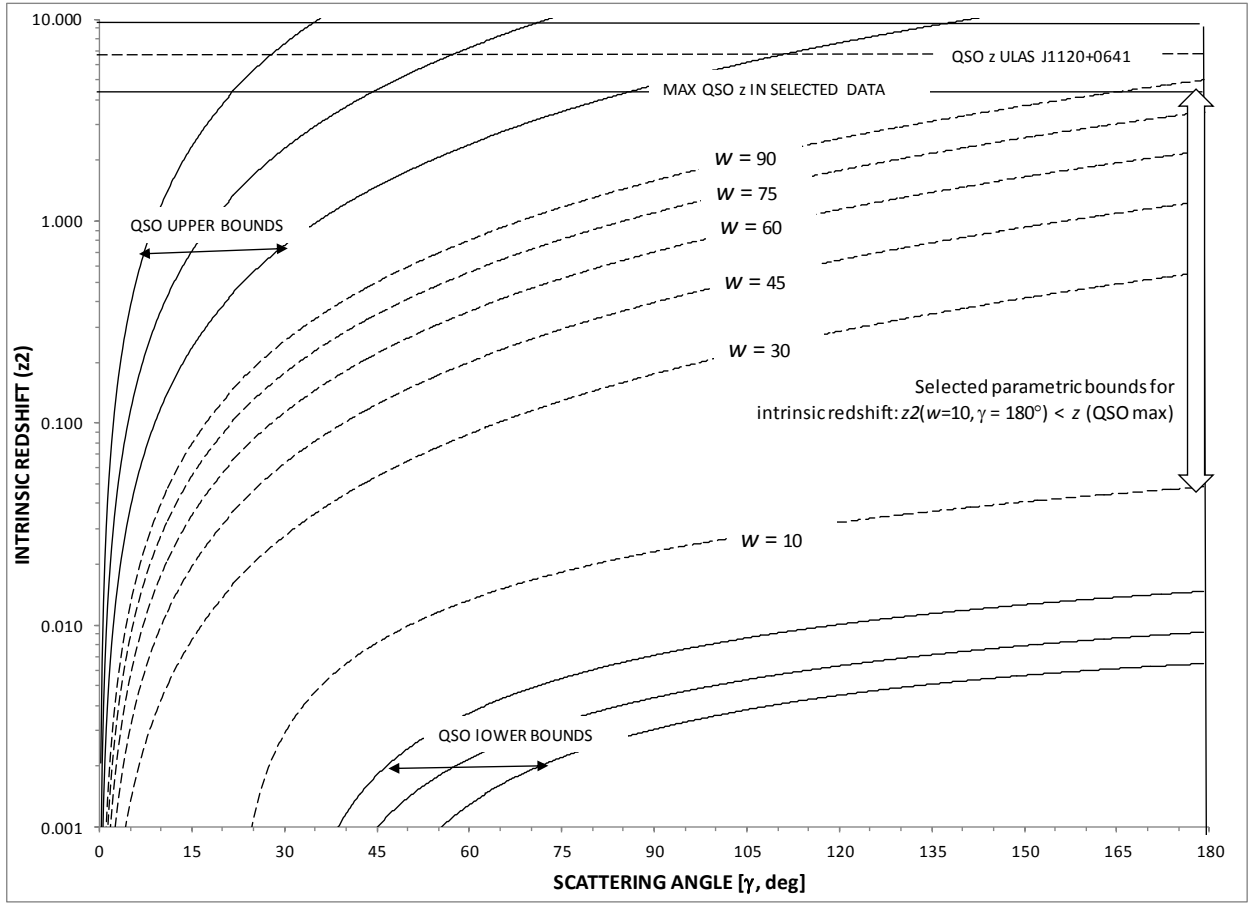

Figure 4. Parametric QSO Intrinsic Redshifts

**A Data Base of Quasars and Galaxies**

A longstanding conjecture, or controversy, in the astronomical literature has been a possible galactic origin of ejected quasars that appear to be associated in the optical field with nearby galaxies. The galaxies have relatively low redshifts, while the nearby quasars have high redshifts. If a physical association exists between them, their redshifts, or distances, would be expected to be closer relative to their apparent cosmological or Hubble ($H_0$) distance ($cz/H_0$). One theory to explain the higher redshifts seen in the quasars is a non-recessional or intrinsic component. Since this possibility is the subject of the present study, a data base[2] of apparently associated galaxies and quasars has been selected (Arp 2003), shown in Figure 5. In the figure quasar redshifts ($z$) are shown as a function of the redshift ($Z$) of their apparently associated galaxies.

As can be seen by inspection in the figure, the galactic redshifts cluster roughly around 0.01, while the quasar redshifts are about two orders of magnitude larger at about 1.0. The dashed line represents equivalence between the quasar and galactic redshifts.

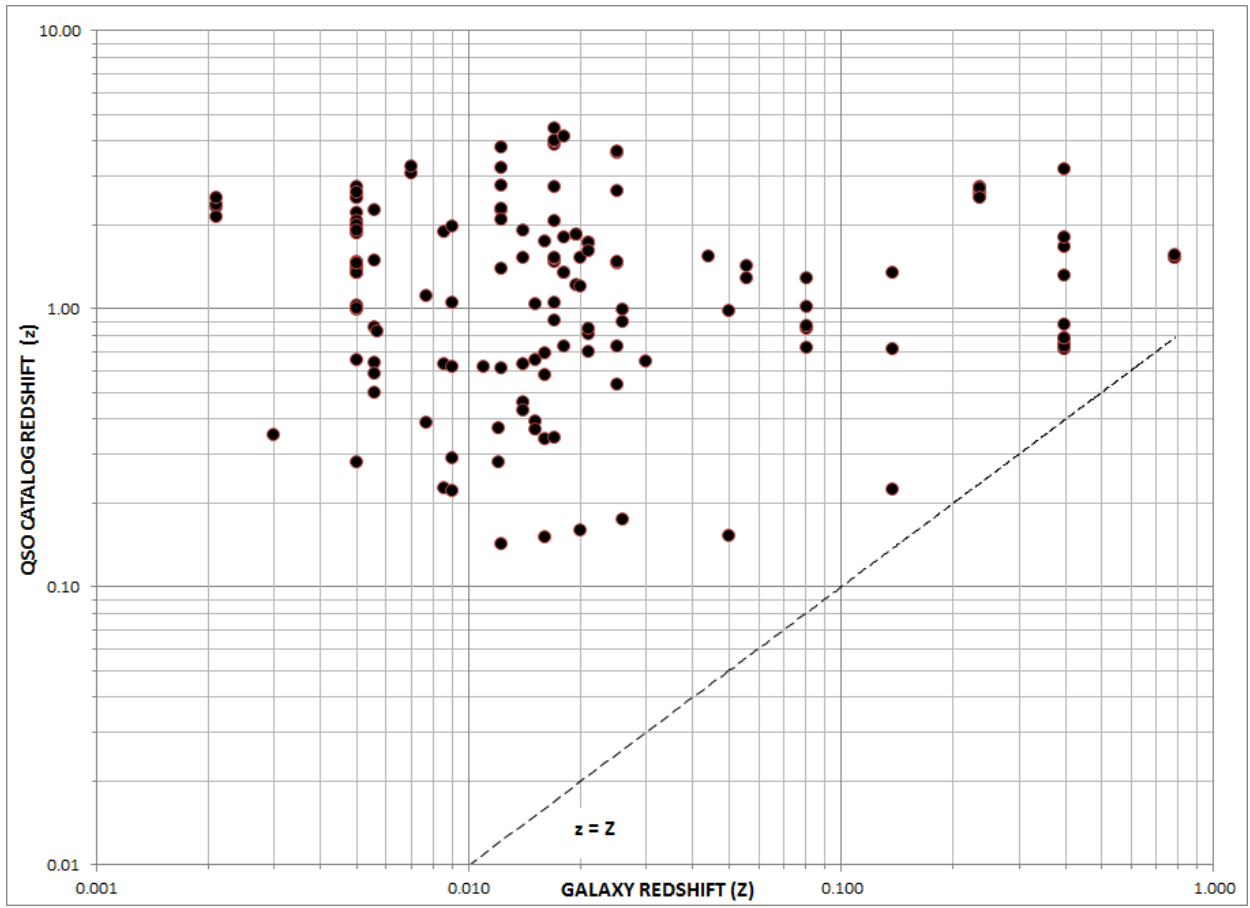

Figure 5. Data Base of Selected Quasar Redshifts vs. Associated Galaxy Redshifts

Given a set of intrinsic redshifts ($z_2$) in Figure 4, and the collection of observed QSO redshifts ($z$) in Figure 5, the recessional redshift component ($z_1$) can be calculated from equation 3 given earlier. Figure 6 shows a set of recessional redshifts based on the bounded data in Figure 4 and the parametric multiplier "$w$". Two additional points are mentioned here: a) in a much earlier study (Hansen 2006) it was shown that if $w \geq 10$ any isotropic galactic ejection would appear as a *redshift*, even if directed at the observer (forward scattering); and b) the ejection velocity ($v_1/c$) exceeds the escape velocity ($v_{esc}/c$) calculated from a data set of 63 galaxies shown in Figure 6.

---

[2] Note: This data base or catalog is a limited sample of approximately 186 objects listed in right ascension, epoch 2000. Figure 5 has been limited by inspection to 150 of those objects that appear proximally associated. It should be emphasized that this data base does not represent the full range and distribution of QSO surveys

Figure 6 also shows that as the observed, or catalog, redshift of a QSO increases, the intrinsic redshift $z_2(w)$ must increase (increasing $w$) as well to exceed the galactic escape velocity if it is indeed an ejection phenomena.

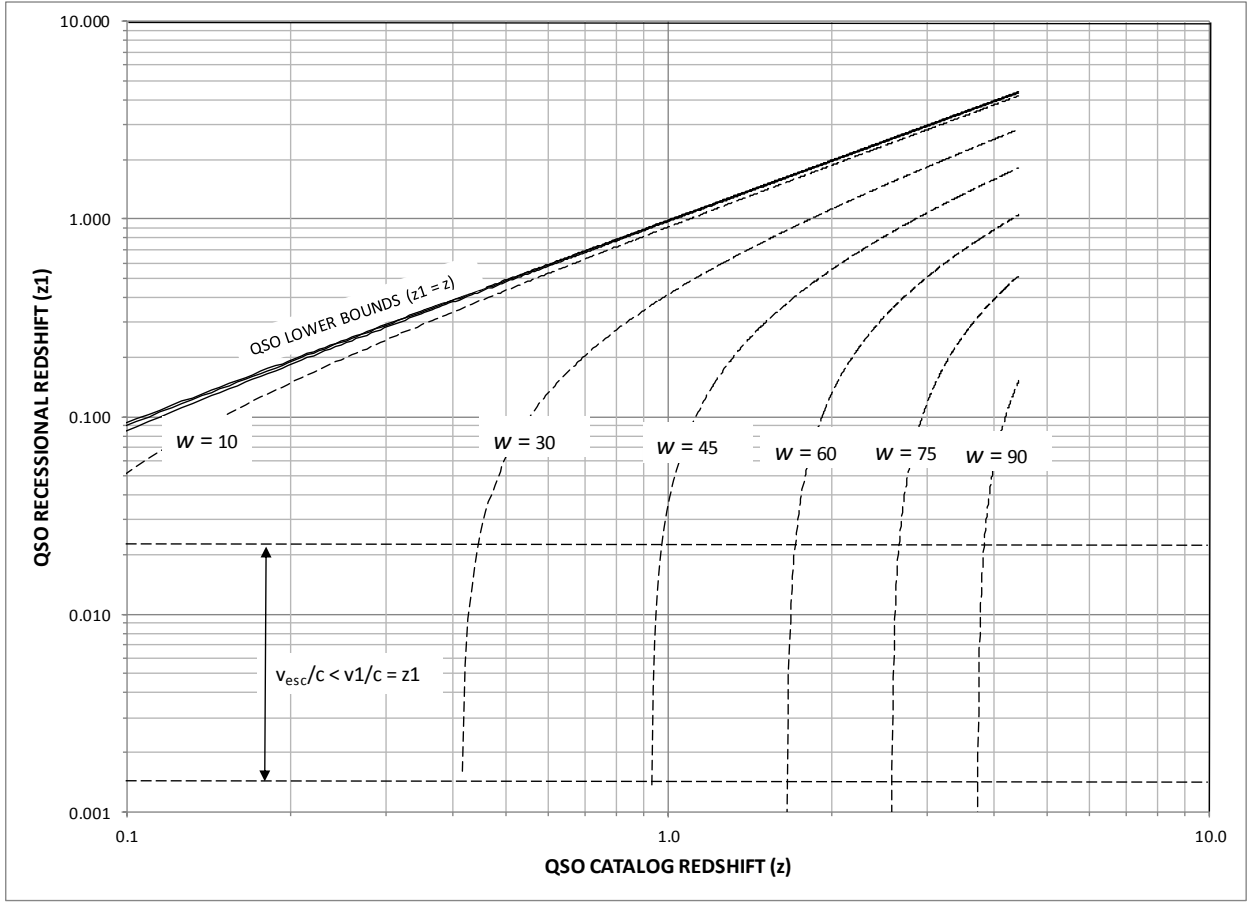

Figure 6. Parametric QSO Recessional Redshifts

**QSO Distance and Luminosity Relations**

The recessional distance of a QSO (using $z_1$) absent the effects of intrinsic redshift ($z_2$) can be found as a *percent* of the apparent, or catalog, redshift as: $z_1/z$ shown as a function of $z$ in Figure 7 for the set of parametric intrinsic redshifts $z_2(w)$: as effects of the intrinsic redshift increase, the QSO distance decreases. This is also illustrated in Figure 8 for two values of the intrinsic redshift parameter $w$ applied to the astronomical QSO catalog redshift data in Figure 5 leaving just the recessional component. Note the increase in proximity or association relative to the galactic distance or redshift ($Z$). However, it should be noted that the most correct comparison from Figure 8 to Figure 5 would employ a transformation of the QSO redshifts or recessional velocities into the reference frame of the apparent associated galaxies. This was not done. The comparison to Figure 5 is only meant to be suggestive of the possible galactic association.

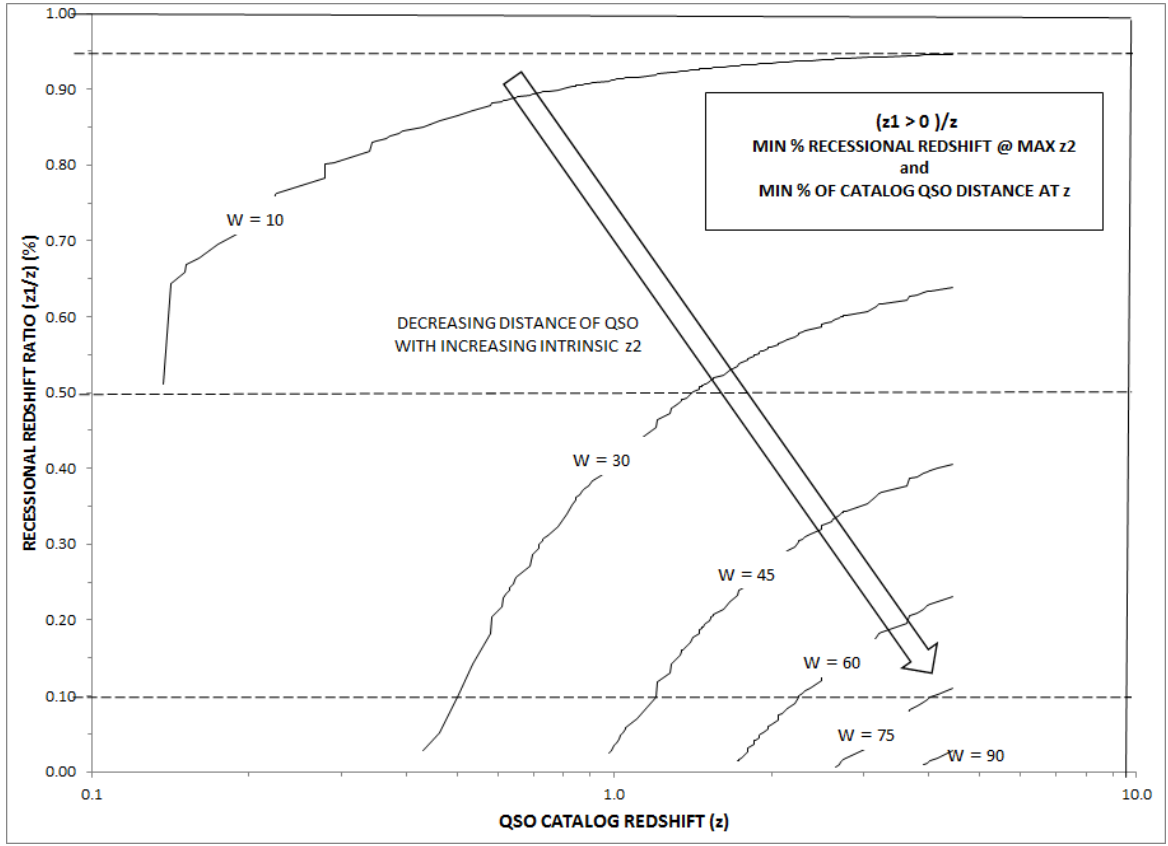

Figure 7. Decreasing QSO Recessional Distances with Intrinsic Redshift Removed

Given a closer recessional distance for the QSO's than the catalog distance as discussed above, the intrinsic luminosity would also decrease. Assuming the QSO power density at the earth is constant, the QSO luminosity ratio, given as a *percent* of the apparent catalog luminosity, can be found from: $(z_1/z)^2$ as shown in Figure 9 as a function of *z*.

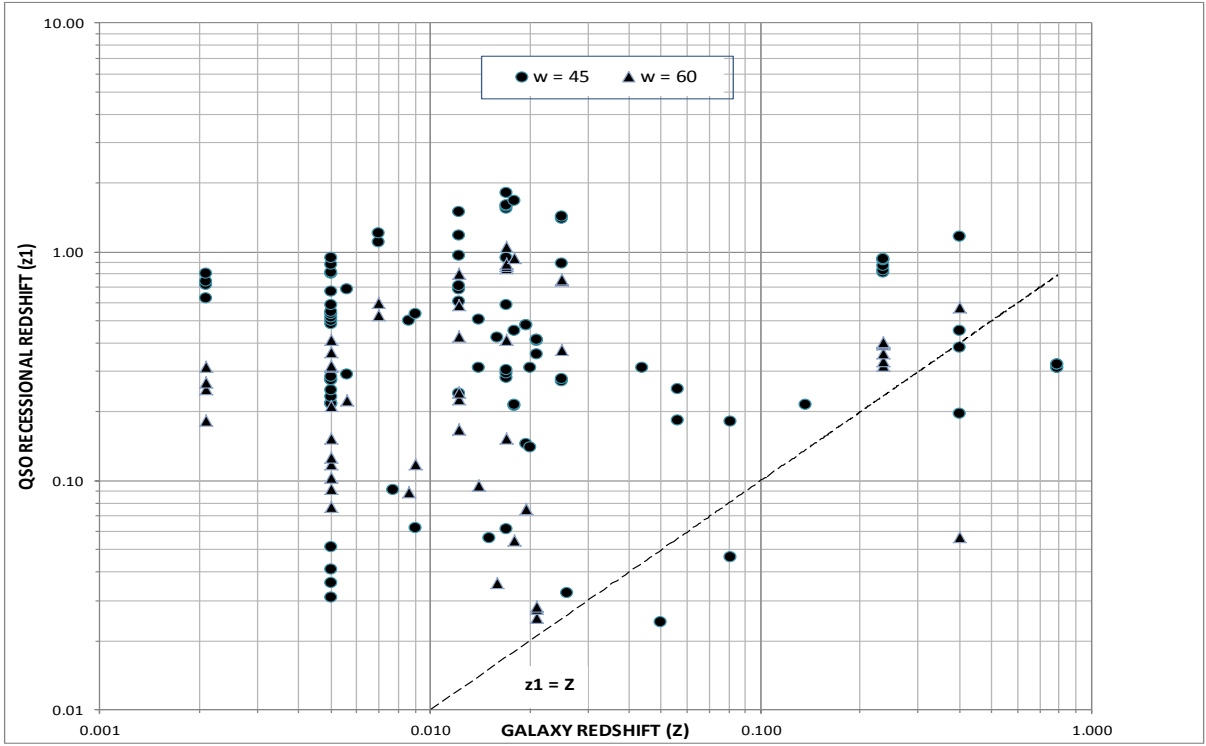

Figure 8. QSO Recessional Distances Relative to Apparent Associated Galaxies

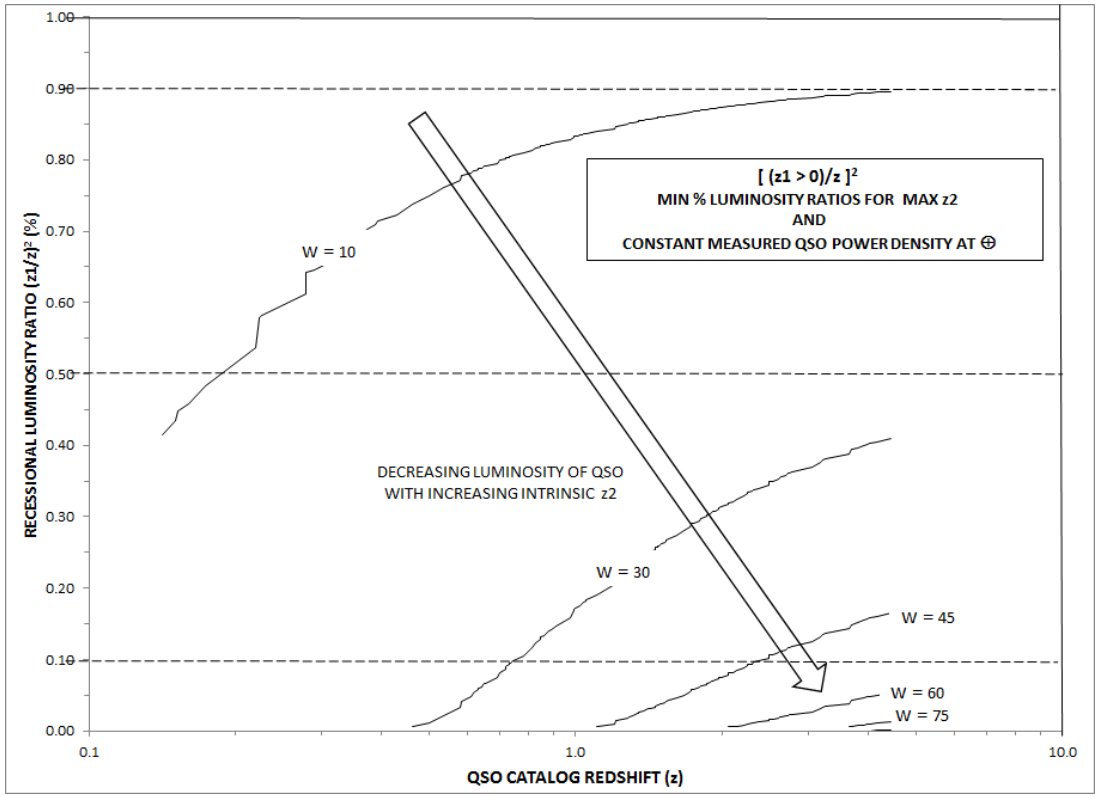

Figure 9. Decreasing QSO Luminosity with Recessional Distance

**Discussion**

This study has focused on the effects of a QSO ambient plasma and the intrinsic redshift it could produce. In the absence of a detailed physical model of the QSO environment it has been necessary to use scaling and parametric techniques. Given the nature of what we can estimate about a QSO's temperature and electron density for example, there should be no controversy about the presence of a plasma component. Exactly how it can effect an intrinsic redshift through correlations in its dielectric susceptibility remains theoretical, but experimentally verified in principal as mentioned earlier. This effect should be taken seriously and investigated further by the astrophysics community! This has not happened despite attempts to draw attention to it since Wolf's work in the 1980's.

The other idea briefly examined in this study was the possibility of high redshift QSO's as ejections from apparent associated lower redshift galaxies. This is a much more tenuous and controversial topic. As discussed, however, the recessional QSO redshifts or distances, absent the intrinsic effect, approach nearer to those of the galaxies in question, and the magnitude of the intrinsic effect ensure no unobserved blueshifts in the case of isotropic ejections.